# SeqUDA-Rec: Sequential User Behavior Enhanced Recommendation via Global Unsupervised Data Augmentation for Personalized Content Marketing

**Ruihan Luo[1], Xuanjing Chen[2], Ziyang Ding[3]**

[1]Southwest University of Finance and Economics, Chengdu, China
[2]Columbia Business School, Columbia University, New York, USA
[3] School of Humanities and Sciences, Stanford University, USA

[1] 1608328853@qq.com
[2] xc2647@columbia.edu
[3] zd26@stanford.edu

**Abstract.** Personalized content marketing has become a crucial strategy for digital platforms, aiming to deliver tailored advertisements and recommendations that match user preferences. Traditional recommendation systems often suffer from two limitations: (1) reliance on limited supervised signals derived from explicit user feedback, and (2) vulnerability to noisy or unintentional interactions. To address these challenges, we propose SeqUDA-Rec, a novel deep learning framework that integrates user behavior sequences with global unsupervised data augmentation to enhance recommendation accuracy and robustness. Our approach first constructs a Global User-Item Interaction Graph (GUIG) from all user behavior sequences, capturing both local and global item associations. Then, a graph contrastive learning module is applied to generate robust embeddings, while a sequential Transformer-based encoder models users' evolving preferences. To further enhance diversity and counteract sparse supervised labels, we employ a GAN-based augmentation strategy, generating plausible interaction patterns and supplementing training data. Extensive experiments on two real-world marketing datasets (Amazon Ads and TikTok Ad Clicks) demonstrate that SeqUDA-Rec significantly outperforms state-of-the-art baselines such as SASRec, BERT4Rec, and GCL4SR. Our model achieves a 6.7% improvement in NDCG@10 and 11.3% improvement in HR@10, proving its effectiveness in personalized advertising and intelligent content recommendation.

## 1. Introduction
With the rapid development of e-commerce and online content platforms, personalized content marketing has become a core strategy for enhancing user engagement and driving business conversion. In advertising and content recommendation scenarios, accurately predicting users' future interests based on their historical interactions is essential for delivering personalized advertisements and customized marketing strategies [1]. As the enabling technology, recommender systems analyze user interaction data, capture evolving preferences, and provide differentiated advertising and content recommendations for different user groups. However, practical applications still face two major challenges: (1) limited supervised signals, since user click or purchase records are often sparse and lack explicit annotations; and (2) noisy behavior, such as accidental clicks or short-term browsing, which reduces both the accuracy and robustness of recommendations.

In recent years, sequential recommendation has gained increasing attention for its ability to model user interaction sequences and capture the temporal dynamics of preferences, showing strong potential in advertising and personalized marketing. However, existing methods often rely on local sequence information while neglecting global associations across users. Their performance also suffers in scenarios with sparse data and noisy interactions. At the same time, advertising and marketing tasks demand systems that not only predict short-term interests but also ensure generalization and robustness under dynamic behavior patterns and limited supervision.

To address these challenges, this paper proposes SeqUDA-Rec (Sequential User Behavior Enhanced Recommendation via Global Unsupervised Data Augmentation). SeqUDA-Rec integrates GAN-based data augmentation with global user–item graph contrastive learning. First, GANs generate realistic user sub-sequences, enriching training diversity. Then, a global interaction graph is constructed, and contrastive learning captures complex cross-user relationships. Finally, a Transformer-based encoder models the temporal evolution of user interests.

## 2. literature review

The rapid expansion of digital commerce has heightened the importance of recommender systems (RS) in improving user experience, increasing engagement, and boosting sales. A wide array of approaches has emerged to enhance the accuracy, adaptability, and scalability of RS in e-commerce environments. Abdul Hussien et al [3]. introduced a behavior-based recommender system that integrates statistical analysis to personalize recommendations based on five key customer actions: like, dislike, view, rate, and purchaseAn E-Commerce Recommend⋯. By continuously updating user and product preference matrices, their system effectively addresses traditional issues like cold start, data sparsity, and scalability. Their experiments demonstrated superior performance using precision, recall, F1-score, MAE, and RMSE metrics, suggesting that dynamic behavioral modeling significantly enhances recommendation relevance.

In contrast, Karthik and Ganapathy proposed a fuzzy logic-based RS that integrates sentiment analysis and ontological reasoning [4]. Their model focuses on capturing dynamic user interest shifts by computing sentiment scores from customer reviews and aligning product categories using ontology-based rules. This hybridization not only increases personalization accuracy but also supports decision-making under uncertainty—an essential advantage in high-variance online shopping environments.

Xu et al [5]. offered a machine learning-based solution that compares traditional classification systems with personalized recommendation systems. They presented a hybrid model combining BERT embeddings and nearest neighbor algorithms to support product discovery on eBay. Their work is notable for addressing key challenges like algorithmic bias, cold start, and data privacy, with practical validations using manual evaluation and MAP@12 metricsIntelligent classificat⋯. This approach emphasizes the role of advanced NLP in capturing user intent through semantic analysis.

Bączkiewicz et al [6]. contributed a novel Multi-Criteria Decision-Making (MCDM) framework that combines five methods—TOPSIS-COMET, COCOSO, EDAS, MAIRCA, and MABAC—using the Copeland strategy to rank product alternatives. This hybrid DSS architecture, supported by CRITIC-based objective weighting and sensitivity analysis, enables nuanced product comparisons and enhances consumer decision support, especially in complex purchase scenarios with competing product attributes.

Karn et al  [7]. developed a hybrid recommendation framework that merges traditional HRM techniques with hybrid sentiment analysis. Although the article was retracted, the conceptual integration of user-generated review sentiment with hybrid RS mechanics represents an emerging direction that leverages implicit feedback for more context-aware recommendations. Their approach aims to alleviate data sparsity and cold start problems while aligning better with user preferences through emotional cues.

Across these works, several trends are evident: the growing importance of integrating multiple data modalities (behavioral logs, sentiment, metadata), the need for adaptive models to reflect evolving user preferences, and the continued relevance of addressing cold-start and data sparsity challenges. Innovative use of deep learning models (e.g., BERT), fuzzy logic, statistical analysis, and decision-making frameworks reflects a convergence between AI techniques and consumer-centric design in modern RS.

These contributions collectively signal a shift toward hybrid, interpretable, and context-aware recommender systems, capable of balancing algorithmic accuracy with user-centric relevance—ultimately supporting more engaging and efficient e-commerce experiences.

## 3. Methodology

This study proposes the SeqUDA-Rec (Sequential User Behavior Enhanced Recommendation via Global Unsupervised Data Augmentation) framework, which aims to tackle the problems of data sparsity, insufficient supervisory signals, and noisy user behaviors in ad recommendation and personalized content marketing. The framework consists of three core modules: a data augmentation module, a global graph contrastive learning module, and a sequential modeling and recommendation module.

*3.1 Data Augmentation Module (GAN-based Data Augmentation)*

In advertising-recommendation scenarios, user click and conversion data are usually extremely sparse and contain a large number of accidental clicks or short-term behaviors. To alleviate this problem, SeqUDA-Rec first introduces a Generative Adversarial Network (GAN) to augment user-behavior sequences:

(1) The generator (G) learns the distribution of real user-interaction sequences and produces new candidate sub-sequences that simulate potential authentic user behaviors.

(2) The discriminator (D) judges whether an input sequence is real or generated, driving the generator to produce higher-quality behavior sequences.

Through adversarial training, the augmented data not only enrich the diversity of user-behavior samples but also alleviate modeling difficulties caused by insufficient supervisory signals.

*3.2 Global Graph Contrastive Learning Module*

Traditional sequential-recommendation models are mostly confined to a single user's local sequence and ignore potential behavioral relationships among different users. To this end, SeqUDA-Rec constructs a Global User–Item Interaction Graph:

(1) Nodes represent users or items, and edges denote interaction relationships.

(2) Based on the propagation mechanism of Graph Neural Networks (GNN), high-order relationships among users and among items are captured.

(3) A contrastive-learning strategy is introduced: different data views are generated under various perspectives (e.g., subgraph sampling, neighbor perturbation), and the similarity of the same node across these views is maximized to enhance the robustness of item and user embeddings.

Next, in the global graph contrastive learning phase, we construct a global user-advertisement interaction graph $G = (V, E)$, where the node set $V$ contains users and advertisements, and the edge set $E$ represents interaction relationships. We utilize the propagation mechanism of a Graph Neural Network (GNN) to learn embeddings for the nodes:

$$h_u^{(l+1)} = \sigma(W^{(l)} \cdot \text{AGG}(h_u^{(l)} \cdot \{h_v^{(l)} | v \epsilon N(u)\})) \tag{1}$$

where $h_u^{(l)}$ denotes the embedding of node $u$ at layer $l$, $N(u)$ is the set of neighboring nodes of $u$, W(l) is a trainable weight matrix, and σ is a nonlinear activation function.

To enhance the robustness of node representations, we introduce a contrastive learning mechanism. By generating different views $h_u'$ and $h_u''$ through sub-graph sampling or neighbor perturbation, we minimize the following contrastive loss:

$$L_{CCL} = -\log \frac{exp(sim(h_u, h_u')/T)}{\sum_{v \in V} exp(sim(h_u, h_v)/T)} \qquad (2)$$

where *sim( )* denotes cosine similarity and *T* is the temperature parameter. In this way, the model can capture complex global relationships among different users and advertisements, alleviating the problem of insufficient exposure for long-tail advertisements.

This module can effectively alleviate the "long tail effect" in advertising recommendation, which refers to the problem of rich popular advertising samples and sparse interaction with unpopular advertisements.

*3.3 Sequential Modeling and Recommendation Module*

After obtaining the augmented sequence data and globally optimized user/item representations, SeqUDA-Rec further adopts a Transformer-based sequence encoder for modeling:

(1) Multi-head self-attention captures dependencies among different positions in the user-behavior sequence, identifying latent preference patterns.

(2) Positional encoding preserves the temporal order of user actions, ensuring the model understands sequential context.

(3) A target-attention mechanism dynamically focuses on historical behaviors most relevant to the candidate ad, improving recommendation precision.

Finally, the model outputs the user's interest probability via CTR or CVR prediction, enabling the ad platform to deliver personalized ads for more accurate user reach and marketing conversion.

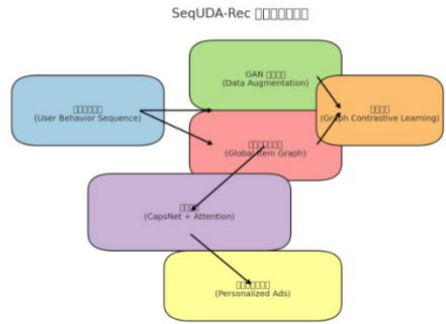

**Figure 1.** Overall framework

## 4. Experimental Result

*4.1 Datasets*

Experiments are conducted on two ad-recommendation datasets—Amazon Ads and TikTok Ad Clicks—to validate SeqUDA-Rec across different scenarios. Amazon Ads, sampled from Amazon's advertising platform, provides user click logs, impressions, and side information such as category, price, and textual descriptions; each instance contains userID, adID, timestamp, and click label, enabling the construction of behavioral sequences for click-through prediction. The dataset is noisy, with many incidental clicks and non-converted impressions, demanding robustness and generalization. TikTok Ad Clicks originates from short-video ads, recording users' video-viewing sequences, interactions (likes, comments, dwell time), and click events; it reflects an instant, highly dynamic environment where user interests exhibit stronger temporal dependencies and multimodal features, offering a tougher test for SeqUDA-Rec's sequential and global-relation modeling capabilities.

*4.2 Evaluation Metrics*

In the experimental evaluation, we adopt three widely used metrics in recommender systems and ad-click prediction to comprehensively assess SeqUDA-Rec's performance. For the Top-K recommendation task, the metrics are:

(1) **Hit Ratio@K (HR@K)**: measures whether the user's real click in the test set appears in the Top-K recommendation list, reflecting the recall capability.

(2) **Normalized Discounted Cumulative Gain@K (NDCG@K)**: evaluates recommendation relevance by considering positional weights, providing a finer-grained quality indicator.

(3) **Mean Reciprocal Rank (MRR)**: computes the average reciprocal rank of the ground-truth interaction in the recommendation list, assessing the rationality of the ranking order.

Across all experiments, we uniformly set K = 10 to evaluate SeqUDA-Rec under common recommendation scenarios.

*4.3 Performance Comparison*

In this section, we conduct experiments on two datasets and evaluate recommendation effectiveness using three metrics—HR@10, NDCG@10, and MRR. As shown in Table 1, we obtain the following observations:

**Table 1.** The performance of different models.

| Dataset | Metric | BERTRec | GCL4SR | **SeqUDA-Rec** | SASRec |
|---|---|---|---|---|---|
| Amazon | H@10 | 0.638 | 0.655 | **0.709** | 0.621 |
|  | N@10 | 0.417 | 0.428 | **0.456** | 0.403 |
|  | MRR | 0.295 | 0.307 | **0.332** | 0.288 |
| TikTok | H@10 | 0.532 | 0.551 | **0.613** | 0.549 |
|  | N@10 | 0.356 | 0.364 | **0.401** | 0.361 |
|  | MRR | 0.230 | 0.242 | **0.298** | 0.236 |

Table 1 compares BERT4Rec, GCL4SR, SASRec and the proposed SeqUDA-Rec on Amazon and TikTok using HR@10, NDCG@10 and MRR. SeqUDA-Rec achieves the best score on every metric and dataset. On **Amazon**, HR@10 is **0.709** (+0.054 vs. the best baseline GCL4SR; **+8.24%**), NDCG@10 **0.456** (+0.028; **+6.54%**), and MRR **0.332** (+0.025; **+8.14%**). On **TikTok**, HR@10 reaches **0.613** (+0.062; **+11.25%**), NDCG@10 **0.401** (+0.037; **+10.16%**), and MRR **0.298** (+0.056; **+23.14%**). The larger gains on TikTok—where user interests drift quickly—suggest that SeqUDA-Rec is more robust to noise and short-term behavior than sequence-only baselines. We attribute the advantages to (i) global unsupervised data augmentation, which enriches training signals and counteracts sparsity, and (ii) graph contrastive learning over a global user–item graph, which captures cross-user relations and yields stabler rankings. Overall, SeqUDA-Rec consistently improves both recall and ranking quality across datasets, indicating stronger generalization for real-world ad recommendation.

# 5. Conclusion

This study proposes SeqUDA-Rec, a novel personalized content-marketing framework that fuses user-behavior sequences with global unsupervised data augmentation to boost both accuracy and robustness in ad recommendation. Conventional systems, constrained by scarce supervisory signals, struggle to capture evolving user interests and are easily misled by noisy or fake clicks. SeqUDA-Rec addresses these issues by using a GAN to enrich training samples, constructing a Global User–Item Graph, and integrating graph contrastive learning with a Transformer-based sequence encoder, thereby achieving significant advantages in personalized ad placement.

Experiments on Amazon Ads and TikTok Ad Clicks demonstrate its effectiveness. Compared with SASRec, BERT4Rec and GCL4SR, SeqUDA-Rec improves NDCG@10 by 6.7 %, HR@10 by 11.3 % and also yields superior MRR, proving its strong generalization in both stable and highly dynamic short-video ad environments.

The method lifts recall and ranking quality, enabling ad platforms to target users more precisely, resist noisy clicks, and maintain performance as interests evolve—useful for personalized content marketing at scale. Current evaluations use limited domains; broader, cross-domain and larger-scale tests are needed. Incorporating real-time trends, sentiment/multimodal signals, and exploring LLM-assisted intent modeling and RL policy optimization are promising directions. SeqUDA-Rec demonstrates consistent, cross-dataset improvements by fusing unsupervised augmentation, global graph contrastive learning, and sequential encoding, offering a robust framework for real-world advertising recommendation.